\documentclass[aps,prc,twocolumn,superscriptaddress,showpacs,longbibliography,lengthcheck]{revtex4-1}

\usepackage{amsmath,amssymb}
\usepackage[dvipdfmx]{graphicx}
\usepackage{color} 
\usepackage{txfonts}

\usepackage{xspace}  
\usepackage{multirow}
\usepackage{dcolumn}

\newcommand{\etal}{\textit{et al.}\xspace}

\begin{document}

%\preprint{}

\title{Triaxial rigidity of $^{166}$Er and its Bohr-model realization} % no punctuation

\newcommand{\acns}{        \affiliation{Center for Nuclear Study, The University of Tokyo, 7-3-1 Hongo, Bunkyo, Tokyo 113-0033, Japan}}
\newcommand{\aut}{         \affiliation{Department of Physics, The University of Tokyo, 7-3-1 Hongo, Bunkyo, Tokyo 113-0033, Japan}}
\newcommand{\ariken}{      \affiliation{RIKEN Nishina Center, 2-1 Hirosawa, Wako, Saitama 351-0198, Japan}}
 
\newcommand{\aemy}{\email{Corresponding author: ytsunoda@cns.s.u-tokyo.ac.jp}}  % author email
\newcommand{\aemt}{\email{Corresponding author: otsuka@phys.s.u-tokyo.ac.jp}}  % author email

\author{Yusuke~Tsunoda}   \aemy \acns   
\author{Takaharu~Otsuka}   \aemt  \aut \ariken

\date{\today}

\begin{abstract}      
The triaxial nature of low-lying rotational bands of $^{166}$Er is presented from the viewpoint of the Bohr Hamiltonian and from that of many-fermion calculations by the Monte Carlo shell model and the constrained Hartree-Fock method with projections.  A recently proposed novel picture of those bands suggests definite triaxial shapes of those bands, in contrast to the traditional view with the prolate ground-state band and the $\gamma$-vibrational excited band.  Excitation level energies and E2 transitions can be described well by the Bohr Hamiltonian and by the many-fermion approaches, where rather rigid triaxiality plays vital roles, although certain fluctuations occur in shell-model wave functions.   Based on the potential energy surfaces with the projections, we show how the triaxial rigidity appears and what the softness of the triaxiality implies.  The excitation to the so-called double $\gamma$-phonon state is discussed briefly.
\end{abstract}

\maketitle

The appearance of the rotational bands is one of the most prominent and widely-seen features of  atomic nuclei at low-excitation energies.  It has been discussed in connection to the deformation of the nuclear shape, since Rainwater \cite{rainwater1950} and Bohr and Mottelson \cite{bohr1952,bohr_mottelson1953,bohr_mottelson_book2}.  As stated already in \cite{aage_bohr_nobel}, a well-accepted picture is that in this appearance, the ground-state (rotational) band is built most likely, if not always, on a prolate deformed ground state, while low-lying rotational bands are formed on top of vibrational excitations from the ground state as an equilibrium.  An example of this picture was presented with $^{166}$Er in \cite{aage_bohr_nobel}, where the second 2$^+$ state was described as a so-called $\gamma$-vibrational state, forming a rotational band on top of it.  

This conventional scenario of the vibrational excitations has been applied to the description of many nuclei \cite{bohr_mottelson_book2}, and has been presented in many textbooks, for instance, \cite{ring_schuck_book}.  In the SU(3) limit of the interacting boson model (IBM)\cite{ibm_su3}, the essentially same underlying picture arises \cite{Warner1982,ibm_cl1,ibm_cl2,ibm_cl3}.  On the other hand, some arguments have been casted over the relevance of this scenario to the structure of real nuclei (see also review papers on the experimental findings \cite{Garrett2001,Sharpey-Schafer2019}).  Quite recently, as described below, another picture of the structure of $^{166}$Er was presented by using the result of the Monte Carlo shell model (MCSM) \cite{mcsm_rev2001,mcsm_rev2012} with a reasonable interaction \cite{Otsuka2019}.  This new picture suggests a different shape of the ground state from the prolate one, and the vibrational aspect is not seen in the lowest excited band.  In this paper, we shall show what shapes emerge for the example of $^{166}$Er, from such quantum many-body approaches as well as from the Bohr Hamiltonian (or
collective Hamiltonian) \cite{bohr1952,bohr_mottelson1953}.

% Fig1: levels
\begin{figure}[tbh]
  \centering
  \includegraphics[width=7.5cm]{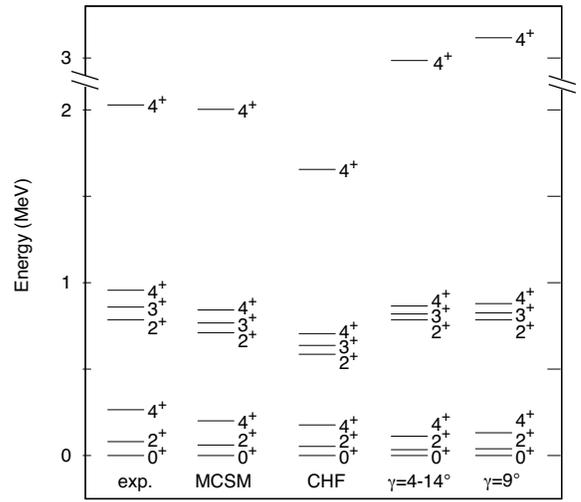}
  \caption{Energy levels of $^{166}$Er nucleus.  Experimental values (exp.) \cite{ensdf} are compared to calculations by Monte Carlo shell model (MCSM) \cite{Otsuka2019}, constrained Hartree-Fock with angular-momentum and parity projections after variation (CHF), Bohr Hamiltonian with a square-well potential ($\gamma$=4-14$^{\circ}$) or a rigid triaxiality ($\gamma$=9$^{\circ}$).
  }   
  \label{fig:level}  
\end{figure}  

The shape of the atomic nucleus can be deformed from a sphere, towards an ellipsoid.  This phenomenon is referred to as the quadrupole (shape) deformation, which indeed comes up with enhanced quadrupole moments in many nuclei.  The quadrupole moment implies, in this work, the quadrupole moment of mass (= proton + neutron) distribution, unless the electric quadrupole moment is explicitly specified.  The quadrupole moment is a rank-two tensor operator.  For a given state, by diagonalizing the matrix of this operator, one obtains three eigensolutions, which correspond to three axes.  The eigenvalues thus obtained are usually translated into the lengths of the ellipsoid of a uniform density, reflecting the density saturation in nuclei.   Thus, the axes and their lengths of the ellipsoid are fixed.  In the case of equal lengths of two axes, which can be the $x$ and $y$ axes without losing the generality, the deformation is called axially symmetric, with the $z$ axis called the symmetry axis.  If the length of such $z$ axis is longer (shorter) than the other two equal ones, the shape is called prolate (oblate).  If the lengths of the three axes are all different from one another, the shape is called triaxial.  In the conventional picture for deformed heavy nuclei, the shape of the ground state is expected overwhelmingly to be prolate (see {\it e.g.} \cite{Kumar1968}), although its origin is an open question.  The triaxial shape is then considered to occur in the ground state of few nuclei.  The nucleus $^{166}$Er is not an exception as stated above \cite{aage_bohr_nobel,bohr_mottelson_book2}.  The triaxial shape can be soft \cite{Wilets1956} or rigid \cite{Davydov1958,Davydov1959}, which implies, respectively, large or small fluctuation in the degree of the triaxiality.  The appearances and features of the rigid triaxiality have been studied, {\it e.g.} in \cite{Hayashi1984,Otsuka1987,Enami2000,Sun2000,Boutachkov2002,Li2010,Gao2015,Chen2017}, as discussed in some detail below. 

\begin{figure}[bt]
  \centering
  \includegraphics[width=8.5cm]{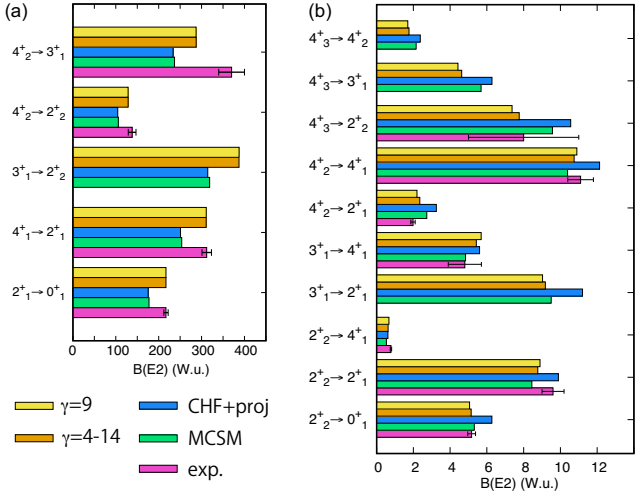}
  \caption{$B(E2)$ values between low-lying states in $^{166}$Er nucleus.  Experimental values (exp.) are included \cite{ensdf}.  Theoretical calculations are labelled similarly to Fig.~\ref{fig:level}.
  }   
  \label{fig:E2}  
\end{figure}

Recently, the configuration interaction (CI) calculation, often called the shell model (calculation) in nuclear physics, has been developed significantly for the study of multinucleon structure of nuclei.  The MCSM is a state-of-the-art large-scale nuclear structure calculation, and was applied to the current topic, the appearance of the rotational bands in $^{166}$Er \cite{Otsuka2019}.  The obtained results display good agreements with experiment for excitation level energies (see Fig.~\ref{fig:level}) as well as for $B(E2)$ values (see Fig.~\ref{fig:E2}) with the standard effective charges $1.5e$ ($0.5e$) for protons (neutrons) \cite{bohr_mottelson_book2}.  The shapes extracted from these results look different from the conventional ones.  We shall look into this intriguing problem.

% Fig2: B(E2) values

The quadrupole(-deformed) shape can be expressed in terms of so-called deformation parameters, $\beta_2$ and $\gamma$.  They imply, respectively, the magnitude of the deformation and the proportion of the ellipsoid axes \cite{bohr_mottelson_book2}.  As visualized in Fig.~3 (a) of \cite{Otsuka2019}, for instance, $\gamma=0^{\circ}$ (60$^{\circ}$) corresponds to a prolate (oblate) shape, and $0^{\circ}<\gamma<60^{\circ}$ implies a triaxial shape.  The deformation parameters are more suitable for the intrinsic state, that is the state in the body-fixed frame.  In the na\"ive modeling, members of a rotational band are obtained by projecting the intrinsic state onto the assigned quantum numbers, {\it e.g.} the total angular momentum $J$, its $z$ component $M$, parity $P$, and some additional ones. 

The potential energy surface (PES) is drawn by the constrained Hartree-Fock (CHF) calculation using the same Hamiltonian for the MCSM calculations.  Here, the intrinsic state is calculated by this CHF calculation with the constraints given by the values of $\beta_2$ and $\gamma$.  Note that they are related to the corresponding quadrupole matrix elements as indicated in \cite{utsuno2015,marsh2018,sels2019}.  Figure~4 (b) of \cite{Otsuka2019} shows the PES thus obtained, exhibiting the minimum around $\beta_2 \sim$ 0.3 and $\gamma \sim 9^{\circ}$.  Note that $\gamma$ is not zero at the minimum, indicative of a triaxial shape. 

The eigenstate of the MCSM is generally expanded by various deformed Slater determinants generated and optimized by the MCSM procedures.  These Slater determinants are called the MCSM basis vectors.  The number of the MCSM basis vectors is usually 50-100, but can be larger if appropriate.  As each MCSM basis vector is a deformed Slater determinant, its quadrupole moments can be calculated, and the corresponding values of $\beta_2$ and $\gamma$ can be obtained.  We can plot individual MCSM basis vectors on the PES according to the $\beta_2$ and $\gamma$ thus evaluated.  Furthermore, the importance of each MCSM basis vector can be visualized by the area of the plot circle, while the importance can be gauged by the overlap probability between the projected and normalized state of the MCSM basis vector and the MCSM eigenstate.  This analysis of the eigenstate properties is called the T-plot \cite{tsunoda2014,otsuka2016}, and has been used in many works, for instance \cite{togashi2016,kremer2016,leoni2017,morales2017,togashi2018,marsh2018,sels2019}.

Figure~4 (c) and (d) of \cite{Otsuka2019} show the T-plot for the 0$^+_{1}$ and 2$^+_{2}$ states, as representative examples.  It was noticed that the T-plot circles show quite similar distributions between these two eigenstates, and appear in the range of $\gamma = 4 - 14^{\circ}$.  Other states, 2$^+_1$, 3$^+_1$ and 4$^+_{1,2}$, exhibit the same feature.  The value of $\beta_2$ of these T-plot circles is confined narrowly to around $\beta_2 = 0.29$.

Once the relevant values of $\beta_2$ and $\gamma$ are clarified, it becomes of great interest to apply the Bohr Hamiltonian \cite{bohr1952,bohr_mottelson1953} to the present study.  The Bohr Hamiltonian has been investigated over decades, as reviewed recently in \cite{Fortunato_2005} and described also in earlier literatures, for instance, \cite{brink_1960,rowe_book,preston_bhaduri_book,casten_book,eisenberg_greiner_book1}.  We first outline the present approach.  The Bohr Hamiltonian is written as
%%%%%%%%%%%%%%%%%%%%%%%%%%%%%%%%%%%%%%%%
\begin{eqnarray}
\label{eq:Bohr}
  H_B &=&-\frac{\hbar^2}{2B} \Bigl[ 
    \frac{1}{\beta_2^4}\frac{\partial}{\partial\beta_2}\beta_2^4\frac{\partial}{\partial\beta_2}
    +\frac{1}{\beta_2^2\sin3\gamma}\frac{\partial}{\partial\gamma}\sin3\gamma\frac{\partial}{\partial\gamma}  \nonumber \\  
  & & \,\,\,\, \,\,\,\,\,\,\,\, \,\,\,\, -\frac{1}{4\beta_2^2}\sum_\kappa\frac{Q_\kappa^2}
       {\sin^2(\gamma-\frac{2}{3}\pi\kappa)} 
  \Bigr]  +V(\beta_2,\gamma),
\end{eqnarray}
%%%%%%%%%%%%%%%%%%%%%%%%%%%%%%%%%%%%%%%%
where $B$ is a parameter, $Q_\kappa\, (\kappa=1,2,3)$ imply angular momentum operators in the body-fixed frame, and $V(\beta_2,\gamma)$ denotes the potential.  For this Bohr Hamiltonian, the Schr\"odinger equation is set as,
%%%%%%%%%%%%%%%%%%%%%%%%%%%%%%%%%%%%%%%%
\begin{equation}
\label{eq:Schr}
H_B \, \Psi(\beta_2,\gamma,\theta_i)=E \, \Psi(\beta_2,\gamma,\theta_i),
\end{equation}
%%%%%%%%%%%%%%%%%%%%%%%%%%%%%%%%%%%%%%%%
where $\theta_i \, (i=1,2,3)$ denote Euler angles for the orientation of a deformed nucleus, and $E$ is the energy eigenvalue.  We assume that $\beta_2$ is a constant.  With this assumption, the above Schr\"odinger equation is reduced to
%%%%%%%%%%%%%%%%%%%%%%%%%%%%%%%%%%%%%%%%
\begin{eqnarray}
\label{eq:gamma}
  & \frac{\hbar^2}{2B\beta_2} \Bigl[
  -\frac{1}{\sin3\gamma}\frac{\partial}{\partial\gamma}\sin3\gamma\frac{\partial}{\partial\gamma}
  +\frac{1}{4}\sum_\kappa\frac{Q_\kappa^2}{\sin^2(\gamma-\frac{2}{3}\pi\kappa)}
  +V_\gamma(\gamma)
  \Bigr] \, \Phi(\gamma,\theta_i) \nonumber \\
 & \,\,\, = E \, \Phi(\gamma,\theta_i), 
\end{eqnarray}
%%%%%%%%%%%%%%%%%%%%%%%%%%%%%%%%%%%%%%%%
where $V_\gamma(\gamma)$ is an appropriate potential.  This equation is solved, by taking the following form, 
%%%%%%%%%%%%%%%%%%%%%%%%%%%%%%%%%%%%%%%%
\begin{equation}
\label{eq:wf}
\Phi(\gamma,\theta_i)=\sum_K g^I_K(\gamma)D^I_{MK}(\theta_i),
\end{equation}
%%%%%%%%%%%%%%%%%%%%%%%%%%%%%%%%%%%%%%%%
where $I$ is the angular momentum of the eigensolution in the laboratory frame, $D^I_{MK}(\theta_i)$ means the usual $D$ function, and $g^I_K(\gamma)$ is the solution or the $K$-component of the wave function to be obtained. Here, $g^I_K(\gamma)$ must obey some symmetry conditions, which ends up with $K$ being even integers with $|K| \le I$ and $g^I_K(\gamma)=(-)^I g^I_{-K}(\gamma)$.  We note that this $K$ corresponds to the usual $K$ quantum number, as it means the $z$ projection of the angular momentum in the body-fixed frame.

We now apply this process to the present problem.  Because the T-plot is concentrated in the range of $\gamma = 4 - 14^{\circ}$ up to 99\%, the first choice of the potential $V_\gamma(\gamma)$ is the square well with a deep attraction for $\gamma = 4 - 14^{\circ}$.  The parameter $B$ is fitted so as to reproduce the observed excitation energy of the second 2$^+$ state. The 4th column of Fig.~\ref{fig:level} shows the excitation energies thus obtained, which agree surprisingly well with the experimental values as well as the MCSM values.  

The E2 transition operator is given, in the lowest order, by
%%%%%%%%%%%%%%%%%%%%%%%%%%%%%%%%%%%%%%%%
\begin{equation}
\label{eq:E2}
T(E2)=t\beta_2\Bigl[D^{(2)}_{\mu,0}\cos\gamma+\frac{1}{\sqrt{2}}
  (D^{(2)}_{\mu,2}+D^{(2)}_{\mu,-2})\sin\gamma\Bigr]
\end{equation}
%%%%%%%%%%%%%%%%%%%%%%%%%%%%%%%%%%%%%%%%
where $t$ is a parameter.  We fit the value of $t\beta_2$ so as to reproduce the observed $B(E2; 2^+_1 \rightarrow 0^+_1)$ value.  Figure~\ref{fig:E2} shows various $B(E2)$ values, including the experimental ones.  One finds remarkably good agreements among those values.
 
The $B(E2)$ values can be put into two groups.  One group depicts large values as shown in Fig.~\ref{fig:E2} (a), whereas the other group small values in Fig.~\ref{fig:E2} (b).  In the conventional picture \cite{aage_bohr_nobel,bohr_mottelson_book2}, this was ascribed to the vibrational origin of the excited side band built on the 2$^+_2$ state: the phonon of the relevant vibration carries $K^P=J^P=2^+$, and this phonon is annihilated in the $2^+_2 \rightarrow 0^+_1$ transition.  Thus, this E2 transition is of vibrational character, which produces a certain collectivity but cannot be as strong as in-band transitions.  

%%%%%%%%%%%%%%%%%%%%%%%%%%%%%%%%%%%%%%%%
% Fig3: wave functions squared of the Bohr model
\begin{figure}[!tb]
  \centering
  \includegraphics[width=7cm]{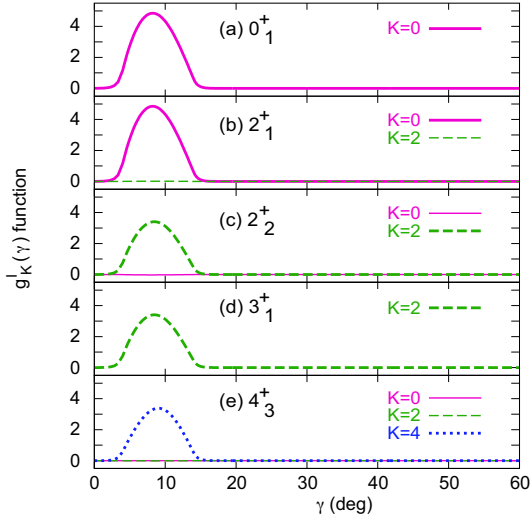}
  \caption{$g^I_K(\gamma)$ functions of various states of $^{166}$Er in the Bohr model for (a) 0$^+_1$, (b) 2$^+_1$, (c) 2$^+_2$, (d) 3$^+_1$, and (e) 4$^+_3$ states, respectively.  The $K^P=0^+$, 2$^+$ and 4$^+$ functions are shown by pink solid, green dashed and blue dotted lines, respectively.  
  }   
  \label{fig:wf}  
\end{figure}  
%%%%%%%%%%%%%%%%%%%%%%%%%%%%%%%%%%%%%%%%

We now turn to the calculation by the Bohr Hamiltonian.  Figure~\ref{fig:wf} shows the function $g^I_K(\gamma)$ for the 0$^+_1$, 2$^+_{1,2}$, 3$^+_1$, and 4$^+_3$ states.  Although the $K$ quantum number is mixed generally for $\gamma \ne 0$, the mixing is weak with the present square well potential.  Thus, these states can be rather well assigned by their primary $K$ values.  As shown in Fig.~\ref{fig:wf}, the $g^I_K(\gamma)$ functions with primary $K$ values exhibit large magnitudes and similar shapes.  The overall magnitude is about $1/\sqrt{2}$ smaller for $K\ne0$, because $K=-I$ component is associated with $K=I$ component.  The $g^I_K(\gamma)$ functions with non-primary $K$ values are negligibly small.

The $g^I_K(\gamma)$ functions of primary $K$ values are peaked around $\gamma = 9^{\circ}$.  We then take a rigid triaxial approximation with this value of $\gamma$: the wave function is a $\delta$ function at this value.  This is nothing but the rigid triaxial model by Davydov and his collaborators \cite{Davydov1958,Davydov1959}.  We can calculate the excitation energies and $B(E2)$ values as shown in the 5th column of Fig.~\ref{fig:level} and in Fig.~\ref{fig:E2} (``$\gamma=9^{\circ}$''), respectively.  One sees almost perfect agreement to the corresponding results of the square well of $\gamma = 4 - 14^{\circ}$.  These quantities are almost linearly dependent on $\gamma$ for $\gamma = 4 - 14^{\circ}$ within the rigid triaxial model.  With the almost symmetric wave function in both sides of $\gamma = 9^{\circ}$, their mean values are close to the corresponding central values, which are nothing but the values given by the Davydov model.  The agreement between the two calculations can thus be understood.  The O(6) case of the IBM \cite{Arima1979} exhibits similar feature with a fully flat potential ($\gamma=0-60 ^{\circ}$) \cite{Otsuka1987}.  

%%%%%%%%%%%%%%%%%%%%%%%%%%%%%%%
% Fig4: PES
\begin{figure}[tb]
  \centering
  \includegraphics[width=6.5cm]{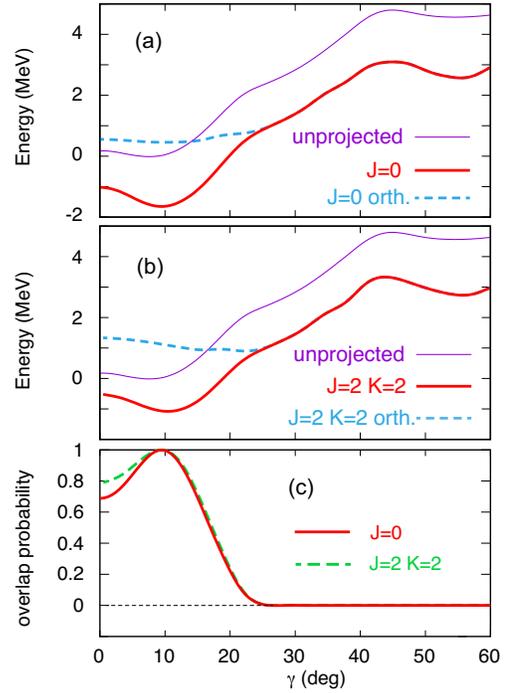}
  \caption{(a,b) Potential energy surfaces (PES) as a function of $\gamma$ with $\beta_2=0.29$.  The unprojected result is shown by purple thin lines.    All values are shown relative to the minimum of this unprojected PES.  The angular-momentum and parity projected results, $\langle\phi(\gamma)|H|\phi(\gamma) \rangle$ in eq.~(\ref{eq:ex}), are shown by red lines (a) for $J^P=0^+$ and (b) for $J^P=K^P=2^+$.  The orthogonalized results, $\langle\eta(\gamma)|H|\eta(\gamma) \rangle$ in eq.~(\ref{eq:ex}), are shown by blue dashed lines.  (c) Overlap probability $|\alpha_{\gamma}|^2$ in eq.~(\ref{eq:wf_orth}).
  }   
  \label{fig:PES}  
\end{figure}  
%%%%%%%%%%%%%%%%%%%%%%%%%%%%%%%%%%%%%%%%

Thus, we conceive the picture that the structure of low-lying states of $^{166}$Er can be described by the simple rigid triaxial shape, but a softer triaxiality, such as $\gamma = 4 - 14^{\circ}$, is equally good, producing almost the same wave functions for various $K$ values as displayed in Fig.~\ref{fig:wf}.  This certainly differs from the conventional picture composed of a prolate ground state and a $\gamma$ vibration on top of this.  We, however, encounter the question as to whether or not the present picture contradicts the shallow PES in the $\gamma$ direction shown in Fig.~4 of \cite{Otsuka2019}.  In order to look into this question, we go back to the shell model study in \cite{Otsuka2019}, and depict the PES in Fig.~\ref{fig:PES} for $\beta_2 = 0.29$ fixed, relative to the lowest energy of the unprojected state.  This is labelled as ``unprojected'' in Fig.~\ref{fig:PES}, because neither angular-momentum nor parity projection is made.  The minimum of this PES is located around $\gamma=10^{\circ}$ away from the prolate point $\gamma = 0^{\circ}$, but the bottom of the PES appears rather shallow.  The structure of $^{166}$Er has been studied by a variety of theoretical works, where the ground state was described as a prolate state and the $K^P=2^+$ band was treated as a $\gamma$ vibrational band (see, for instance, \cite{Nesterenko1993,Garcia-Ramos2000a,Garcia-Ramos2000b,Boutachkov2002,Delaroche2010,Li2010,Bonatsos2011,Inci2011,Nesterenko2016,Chen2017}). It has remained, however, a challenge to achieve a precise microscopic description of both the excitation energies and the B(E2) values involving the $K^P=2^+$ band. We mention that among such works, the PES minimum seems to deviate from $\gamma=0$ to $\gamma \sim 5^{\circ}$ in \cite{Li2010} and to $\gamma \sim 9^{\circ}$ in \cite{Chen2017}.  The minima, however, appeared to be too shallow (by $\lesssim$ 0.2 MeV) to drive the ground state off the prolate shape.
The angular-momentum unprojected and projected calculations were compared in \cite{Hayashi1984} (prior to \cite{Li2010,Chen2017}) for two nuclei $^{168}$Er and $^{188}$Os, of which the former is close to $^{166}$Er.  For $^{168}$Er, the unprojected and projected PESs were rather similar to each other with shallow minima like the cases mentioned above \cite{Li2010,Chen2017}, suggesting an axially symmetric ({\it i.e.} prolate) nucleus  \cite{Hayashi1984}.   For $^{188}$Os, in contrast, the projected calculation produced a pronounced minimum at $\beta_2\sim$0.2 and $\gamma\sim$30$^{\circ}$ contrary to the unprojected calculation \cite{Hayashi1984}.  Because of significant structure difference between $^{188}$Os and  $^{166}$Er, the findings on $^{188}$Os do not seem to be relevant to the present work. 

We now perform the angular-momentum and parity projection on the CHF states, {\it i.e.} a variation before projection (VBP) calculation. We keep $\beta_2 = 0.29$, while $\gamma$ is varied.  Figure~\ref{fig:PES} (a) displays the projected PES thus obtained for $J^P=0^+$: the curve becomes steeper around the minimum compared to the unprojected curve. A similar lowering is seen also for a local minimum near $\gamma = 55^{\circ}$.    

The lowering of the PES by 0.63 MeV occurs at the minimum point, denoted by $\gamma_0$ hereafter.  Its actual value is $\gamma_0 \sim9.5^{\circ}$.  As this lowering is substantially larger than in other works mentioned above, it makes the minimum more profound (see Fig.~\ref{fig:PES} (a)), being consistent with the distinct triaxiality in the present work.  The lowering is natural as the unprojected result includes effects of higher lying states including $K \ne 0$.  This lowering still appears for $J^P=K^P=2^+$ as shown in panel (b), but becomes slightly weaker.  Note that for $J^P=0^+$, only $K^P=0^+$ is possible, though not explicitly mentioned.  The lowering due to the angular-momentum projection was shown, in \cite{Hayashi1984}, to be large for strong $K$ mixings, as in $^{188}$Os, but small for weak $K$ mixings, as in $^{168}$Er.  We point out that in the present work, a strong lowering occurs with $K$ quantum numbers practically conserved, consistently with smaller $\gamma$ values.  The additional lowering in the present work is largely due to the monopole-quadrupole interplay \cite{Otsuka2019}.

We now discuss the relation between the lowering around $\gamma_0$ and the rigid triaxiality.  For this purpose, we introduce the angular-momentum and parity projected state, $\phi(\gamma)$, for a given value of $\gamma$, and expand it by the state of $\phi(\gamma_0)$ and the remaining orthogonal component $\eta(\gamma)$ as
%%%%%%%%%%%%%%%%%%%%%%%%%%%%%%%%%%%%%%%%
\begin{equation}
\label{eq:wf_orth}
\phi(\gamma)=\alpha_{\gamma} \, \phi(\gamma_0) + \sqrt{1-\alpha_{\gamma}^2} \,  \eta(\gamma),
\end{equation}
%%%%%%%%%%%%%%%%%%%%%%%%%%%%%%%%%%%%%%%%
where the amplitude is chosen to be a real number with $\alpha_{\gamma} = \langle \phi(\gamma_0) \,|\, \phi(\gamma) \rangle$, by choosing the phase of $\phi(\gamma)$.  Note that $\phi(\gamma)$ and $\eta(\gamma)$ are normalized.  The overlap probability $|\alpha_{\gamma}|^2$ is depicted in Fig.~\ref{fig:PES} (c) for $J^P=0^+$ and $J^P=K^P=2^+$. 

We now discuss how the expectation value of the Hamiltonian with respect to $\phi(\gamma)$ varies as a function of $\gamma$.  It can be written as
%%%%%%%%%%%%%%%%%%%%%%%%%%%%%%%%%%%%%%%%
\begin{eqnarray}
\label{eq:ex}
&\langle \phi(\gamma)|H|\phi(\gamma) \rangle &= \alpha_{\gamma}^2 \, \langle \phi(\gamma_0)|H|\phi(\gamma_0) \rangle
+\, (1-\alpha_{\gamma}^2) \langle \eta(\gamma)|H|\eta(\gamma) \rangle  \nonumber \\
& & \,+\, 2 \alpha_{\gamma} \sqrt{1-\alpha_{\gamma}^2}\mathrm{Re} \langle \eta(\gamma)|H|  
 \phi(\gamma_0) \rangle.  
\end{eqnarray}
%%%%%%%%%%%%%%%%%%%%%%%%%%%%%%%%%%%%%%%%
Because $\phi(\gamma_0)$ is close to an eigenstate and $\eta(\gamma)$ is orthogonal to it, the last term on the right-hand side can be neglected.  Figure~\ref{fig:PES} (a) indicates $\langle \eta(\gamma)|H|\eta(\gamma) \rangle$ for $J^P=0^+$.  It stays rather constant as $\gamma$ increases from $0^{\circ}$, and starts to ascend after $20^{\circ}$.  This implies that in the second term on the right-hand side of eq.~(\ref{eq:ex}), $\langle \eta(\gamma)|H|\eta(\gamma) \rangle$ is approximated by a constant $\langle \eta(\gamma_0)|H|\eta(\gamma_0) \rangle$ for $0^{\circ}<\gamma<20^{\circ}$.  We thus obtain
%%%%%%%%%%%%%%%%%%%%%%%%%%%%%%%%%%%%%%%%
\begin{eqnarray}
\label{eq:ex1}
&\langle \phi(\gamma)|H|\phi(\gamma) \rangle &\approx  \langle \eta(\gamma_0)|H|\eta(\gamma_0) \rangle \nonumber \\
& &  +\, \alpha_{\gamma}^2 
     \Bigl\{ \langle \phi(\gamma_0)|H|\phi(\gamma_0) \rangle 
                       - \langle \eta(\gamma_0)|H|\eta(\gamma_0) \rangle  \Bigr\},
\end{eqnarray}
%%%%%%%%%%%%%%%%%%%%%%%%%%%%%%%%%%%%%%%%
where $\langle \eta(\gamma_0)|H|\eta(\gamma_0) \rangle$ implies the limiting value with $\gamma \rightarrow \gamma_0$.  As Fig.~\ref{fig:PES} (c) shows the values of $\alpha_{\gamma}^2$, the expectation value $\langle \phi(\gamma)|H|\phi(\gamma) \rangle$ should exhibit a behavior similar to the corresponding curve in Fig.~\ref{fig:PES} (c) with the opposite sign.  The quantity on the left-hand side of eq.~(\ref{eq:ex1}) is reproduced remarkably well by the right-hand side, for instance, within 0.05 MeV for $4 ^{\circ}<\gamma<14 ^{\circ}$.  It is now clear that the present lowering of the $J^P=0^+$ energy seen between  $\gamma = 0^{\circ}$ and $\sim 20^{\circ}$ is predominantly due to the varying large mixing of $\phi(\gamma_0)$.  Thus, the underlying feature of the projected PES points to the crucial role of the triaxial intrinsic state at $\gamma_0$. Panel (b) shows a similar feature for $J^P=K^P=2^+$. 

The present analysis indicates that the triaxial intrinsic state of $\gamma=\gamma_0$ is the origin of the additional binding energy due to the triaxiality in $^{166}$Er, and this effect is common to low-lying states, including the $K^P$=2$^+$ band members, {\it etc}.  A certain softness usually emerges around $\gamma=\gamma_0$, once various correlations are included, for instance in the MCSM calculation.  The intrinsic state of $\gamma=\gamma_0$ remains a major component of such $\gamma$-soft wave functions spreading around $\gamma=\gamma_0$, as discussed above.  Thus, the triaxial rigidity shows two facets: rigid intrinsic state and its characteristic inclusion in $\gamma$-soft wave functions.  Because of this, E2 matrix elements can look like those of the Davydov model \cite{Davydov1958,Davydov1959}, even if the states are $\gamma$ soft to a certain extent.  We note that the contributions from $\eta(\gamma<\gamma_0)$'s  and those from $\eta(\gamma>\gamma_0)$'s tend to cancel each other with $\gamma$ around $\gamma_0$, which definitely enhances this feature.

A rigid-triaxial intrinsic state was assumed in the triaxial projected shell model, although  the terminology of the $\gamma$ vibration was used \cite{Sun2000,Boutachkov2002}.   Thus, one finds an aspect somewhat in common with the present work.  The fitted value $\gamma \sim26^{\circ}$ \cite{Sun2000}, however, substantially differs from the present values (see Figs.~\ref{fig:level} and \ref{fig:E2}).  The calculation may become closer to the present ones, by including 
monopole effect, $\gamma$-softness, {\it etc}.   

Figure~\ref{fig:level} includes the $4^+_3$ states at higher excitation energies.  The experimentally observed state \cite{Fahlander1996} was considered to be $K^P$=4$^+$ two-phonon state.  They are also $K^P$=4$^+$ state in the present MCSM and CHF calculations: the one in the CHF is obtained by projecting the same intrinsic state as the one for the ground and low-lying states, and the MCSM shows a similar feature through T-plot.  They do not correspond to a double phonon excitation.  As shown in Fig.~\ref{fig:wf}, the present calculations by the Bohr Hamiltonian predict it as a member of the group of triaxial states.  Figure~\ref{fig:PES} indicates that the projected PES for $J^P=K^P=2^+$ is slightly more $\gamma$-soft than the PES for $J^P=0^+$.  The projected PES for $J^P=K^P=4^+$ is somewhat softer, which may explain why the excitation energies in the MCSM and CHF are lower than those by the Bohr Hamiltonian where this property is missing.  A triaxial state analogous to the present $K^P$=4$^+$ state is found in the triaxial projected shell model of \cite{Sun2000}.

In summary, we discussed the underlying structure of the rotational bands of $^{166}$Er.  Whereas the conventional picture assumes the prolate ground state and the $\gamma$-vibrational $K^P=2^+$ band, these low-lying states of $^{166}$Er can be understood as a consequence of the triaxial shape.  Although its rigidity may need further theoretical and experimental confirmations, the shell model results combined with the projected CHF calculation suggest that there is a triaxial rigidity which governs the projected PES and crucially determines E2 observables.  We, however, note certain softness seen in the shell-model wave function.  The Bohr Hamiltonian does not distinguish between the square-well and rigid-triaxial models, as it produces similar results from these two with adequate $\gamma$ values.  Thus, the lowest two bands in $^{166}$Er can be considered to be consequences of the common triaxiality with rigidity.  The origin of this triaxiality lies in the nucleon-nucleon interaction as emphasized as a consequence of the self-organization due to the monopole-quadrupole interplay in \cite{Otsuka2019}, and is not addressed in this paper.  The present picture is not limited to $^{166}$Er, and certainly remains to be explored further for more nuclei, and/or in more states particularly higher $J$'s and excitation energies.  

%%%%%%%%%%%%%%%%%%%%%%%%%%%%%%%%%%%%%%%%%%%%%%

%%%%%%%%%%%%%%%%%%%%%%%%%%%%%%%

\section*{Acknowledgements}
The authors acknowledge the partial support from MEXT as ``Priority Issue on post-K computer'' (Elucidation of the Fundamental Laws and Evolution of the Universe) (hp160211, hp170230, hp180179, hp190160), the partial support from MEXT as ``Program for Promoting Researches on the Supercomputer Fugaku'' (Simulation for basic science: from fundamental laws of particles to creation of nuclei) (hp200130) and JICFuS.  This work was supported in part by MEXT KAKENHI Grant No. JP19H05145.  T.O. acknowledges Prof. P. Van Duppen for valuable comments.  The authors thank Prof. N. Shimizu and Dr. Y. Utsuno for valuable helps.  

\end{document}